\documentclass[journal,twoside,web]{ieeecolor2}
\usepackage{generic}
\usepackage{cite}
\usepackage{amsmath,amssymb,amsfonts}
\usepackage{algorithmic}
\usepackage{graphicx}
\usepackage{textcomp}
\usepackage{caption}
\usepackage{enumerate}
\usepackage[T1]{fontenc}
\usepackage[font=small,labelfont=bf,tableposition=top]{caption}
\usepackage{afterpage}
\def\BibTeX{{\rm B\kern-.05em{\sc i\kern-.025em b}\kern-.08em
    T\kern-.1667em\lower.7ex\hbox{E}\kern-.125emX}}
\begin{document}
\title{Heart Rate and Body Temperature Relationship in Children Admitted to PICU - A Machine Learning Approach}
\author{Emilie Lu, Thanh-Dung Le, \IEEEmembership{Member, IEEE}, \\ Philippe Jouvet M.D., Ph.D., and Rita Noumeir Ph.D.,\IEEEmembership{Member, IEEE}
\thanks{This work was supported in part by the Natural Sciences and Engineering Research Council (NSERC), in part by the Fonds de la recherche en sante du Quebec (FRQS).}
\thanks{Émilie Lu is with the Biomedical Information Processing Lab, École de technologie supérieure, Montréal, Québec, Canada (email:emilie-sokuntheavy.lu.1@ens.etsmtl.ca). }
\thanks{Thanh-Dung Le is with the Biomedical Information Processing Lab, École de technologie supérieure, Montréal, Québec, Canada (email:thanh-dung.le.1@ens.etsmtl.ca). }
\thanks{Philippe Jouvet is with the CHU Sainte-Justine Research Center, CHU Sainte-Justine Hospital, University of Montreal, Montréal, Québec, Canada. (email: philippe.jouvet.med@ssss.gouv.qc.ca) }
\thanks{Rita Noumeir is with the Biomedical Information Processing Lab, École de technologie supérieure, Montréal, Québec, Canada. (email: rita.noumeir@etsmtl.ca) }}

\maketitle

\begin{abstract}
Vital signs have been essential clinical measures. Among these, body temperature (BT) and heart rate (HR) are particularly significant, and numerous studies explored their association in hospitalized adults and children. However, a lack of in-depth research persists in children admitted to the pediatric intensive care unit (PICU) despite their critical condition requiring particular attention. \textit{Objective:} In this study, we explore the relationship between HR and BT in children from 0 to 18 years old admitted to the PICU of CHU Sainte-Justine (CHUSJ) Hospital. \textit{Methods:} We applied Machine learning (ML) techniques to unravel subtle patterns and dependencies within our dataset to achieve this objective. Each algorithm undergoes meticulous hyperparameter tuning to optimize the model performance. \textit{Results:} On a large database of 4006 children admitted in the PICU, our findings align with prior research, revealing a consistent trend of decreasing HR with increasing patient age, confirming the inverse correlation. Furthermore, a thorough analysis identifies Gradient Boosting Machines (GBM) implemented with Quantile regression (QR) as the most fitting model, effectively capturing the non-linear relationship between HR, BT, and age. Through testing the HR prediction model based on age and BT, the predictive model between the 5th and 95th percentiles accurately demonstrates the declining trend of HR with age, while HR increases with BT. Based on that, we have developed a user-friendly interface tailored to generate HR predictions at different percentiles based on three key input parameters: current HR, current BT, and patient's age. The resulting output lets caregivers quickly determine whether a patient's HR falls within or outside the normal range, facilitating informed clinical decision-making. Thus, our results challenge previous studies' presumed direct linear association between HR and BT. \textit{Conclusion:} This study contributes to comprehending the non-linear dynamics between HR, BT, and age in critically ill children with the chosen ML model but also challenges established assumptions about the linear relationship between HR and BT. \textit{Significance:} These findings emphasize the importance of reconsidering traditional assumptions in clinical contexts, potentially changing approaches to understanding physiological indicators, and providing new perspectives for future investigations.
\end{abstract}


\section{Introduction}
\label{sec:introduction}
Pediatric Intensive Care Unit (PICU) patients’ health requires particular attention and ongoing monitoring \cite{slusher2018pediatric}. Due to their severe conditions or illnesses, these children, particularly those aged 0 to 18 years old, have distinct physiological characteristics requiring special medical evaluation and care \cite{heneghan2019characteristics}. In the PICU of CHUSJ, caregivers often use the Clinical decision support systems (CDSS) in their practice to improve the quality of patient care and outcome \cite{chusj_sip_ia}. In this high-risk environment, healthcare practitioners rely on vital sign monitoring as fundamental indicators of patients’ health across all age groups during their hospitalization \cite{brekke2019value}. These indicators assist in making clinical decisions and facilitate prompt interventions if necessary. Among these vital signs, heart rate (HR) and body temperature (BT) are crucial parameters providing essential information about a child’s health, as their relationship can be influenced by age, agitation, stress, infection, shock, physiological distress, or underlying illness \cite{jensen2015relationship,davies2009relationship}. Consequently, HR and BT offer invaluable insights for guiding physicians in clinical decision-making \cite{liu2019early}.  

Firstly, HR is fundamental in assessing a person's overall cardiac health \cite{nelson2020guidelines}. This metric varies depending on the child's age, as highlighted by Pediatric Advanced Life Support (PALS) data, where awake patients' reference HR values are age-dependent \cite{davies2009relationship,fleming2011normal}. Newborns under 28 days of age typically have an HR ranging from 85 to 205 beats per minute (bpm), while infants and toddlers aged 1 to 24 months fall between 100 and 190 bpm. In children from 2 to 10 years old, the normal HR ranges between 60 and 140 bpm. Adolescents between 10 and 18 years old usually maintain an HR averaging between 60 and 100 bpm. Monitoring patients’ HR can be an early indicator of various medical conditions \cite{daymont2015heart}. Deviation from the normal HR range can signal underlying health conditions or potential complications. For example, bradycardia, an abnormally low HR, leads to reduced blood flow to organs, leading to dizziness, fatigue, shortness of breath, loss of consciousness, and signs of heart failure \cite{sidhu2020evaluating}. Conversely, tachycardia, i.e., high HR, can indicate a severe medical condition, such as shock \cite{heal2022association}.

Additionally, BT regulation is vital for assessing an individual’s overall health and reflecting the body's heat production and balance, which are crucial for optimal physiological functions. The regulation of BT is managed by the hypothalamus, a central region within the brain that acts as the body's temperature control center. This process, called thermoregulation, aims to adjust BT based on signals perceived by thermoreceptors in blood and skin. The primary goal is maintaining a normal BT around $37 \pm 0,5$°C, varying slightly in infants and young children. The hypothalamus plays a crucial role in initiating necessary adjustments in temperature fluctuations, whether a decrease or an increase. Extreme temperatures beyond normal ranges indicate health issues: low temperatures (hypothermia) suggest cold responses or medical problems that can potentially lead to severe complications. Elevated temperatures (hyperthermia), triggered by various external factors, such as infection, environmental temperature, or disruptions in the body's thermoregulation system, often signal illness. Consistent temperature checks are important in early illness detection, enabling timely medical interventions. This practice is essential for taking precautionary measures to avert physiological dysfunction and protect organs' structure and function from damage \cite{walker1990clinical}. 

A distinct pattern is observed in the pediatric population from 0 to 18 years, where HR tends to decrease with age after the initial month. Notably, there is an observed tendency for HR to rise during the first month following birth, reaching its peak before gradually decreasing \cite{fleming2011normal}. Insights from studies on hospitalized children in primary care and emergency departments consistently show that BT is an independent factor influencing HR, resulting in an average increase of around 10 bpm per degree centigrade \cite{daymont2015heart,davies2009relationship}. More specifically, the studies \cite{thompson2009deriving} and \cite{heal2022association} highlighted an age-specific influence of BT on HR. Every 1°C BT increase in children with acute infections correlates with an HR rise ranging from 9.9 to 14.1 bpm \cite{thompson2009deriving}. In contrast, in children aged 0-16 years old attending urgent and emergency care, the influence of a $1^{\circ}$C increase in HR shows a range of 8.7 to 13.7 bpm, with the observed average being 12.3 bpm \cite{heal2022association}. 

In adults admitted to the emergency department, it has been observed that a $1^{\circ}$C increase in BT leads to an approximately 7 bpm increase in HR \cite{kirschen2020relationship}. In contrast, among acute patients within the emergency department, the results indicate that the group of patients with a high BT, i.e., greater than $37.2^{\circ}$C, exhibited the most significant changes in HR, with an increase of $7.4 \pm 0.9$ bpm per degree \cite{jensen2015relationship}.  In critically ill adults admitted to the Intensive Care Unit (ICU), it was observed that within the temperature range of $32.0^{\circ}$ to $42.0^{\circ}$C, a linear correlation existed between BT and HR. Specifically, for every $1^{\circ}$C increase within this range, the HR showed a linear increase of $8.35 \pm 0.50$ bpm (p < 0.0001) \cite{broman2021relationship}.


Regarding the mentioned findings above, the diverse techniques employed to model the relationship between HR and BT in children and adults reflect the complexity of this physiological association.  In studies with children, approaches ranged from percentile curves using advanced statistical distributions \cite{daymont2015heart}  to quantile regression with polynomial equations \cite{davies2009relationship}, centile charts \cite{thompson2009deriving}, and multiple linear regression models \cite{heal2022association}. Similarly, in adult studies, techniques such as linear regression \cite{kirschen2020relationship}, linear and multiple variable regression analyses \cite{jensen2015relationship}, and multiple linear regression with various covariates  \cite{broman2021relationship} demonstrated a multifaceted understanding of the HR-BT relationship. These methods illustrate the diverse approaches to understanding the HR-BT relationship in different age groups. The findings highlight the need for age-specific considerations, centile charts, and flexible modeling approaches to improve clinical assessments and guide medical interventions. However, the focus on critically ill children remains significantly limited despite the complexity of their medical conditions. Moreover, acknowledging the existing limitations in which traditional methods may not have captured the complexity of the association, the utilization of Machine Learning (ML) and Deep Learning (DL) techniques would enable a more in-depth analysis to accurately capture the intricate patterns between the two vital signs in critically ill children.

\section{Materials and Methods}
This study was conducted following ethical approval from the research ethics board at CHU Sainte-Justine (CHUSJ). The approved protocol number is 2023-5201. Figure \ref{fig:flowchart} represents an overview of the workflow for this study; there are four main steps, including data collection, data preprocessing, ML modeling, and evaluation of the expected output. Those steps are discussed in detail as follows:

\begin{figure}[h]
    \centering
    \includegraphics[width=0.85\linewidth]{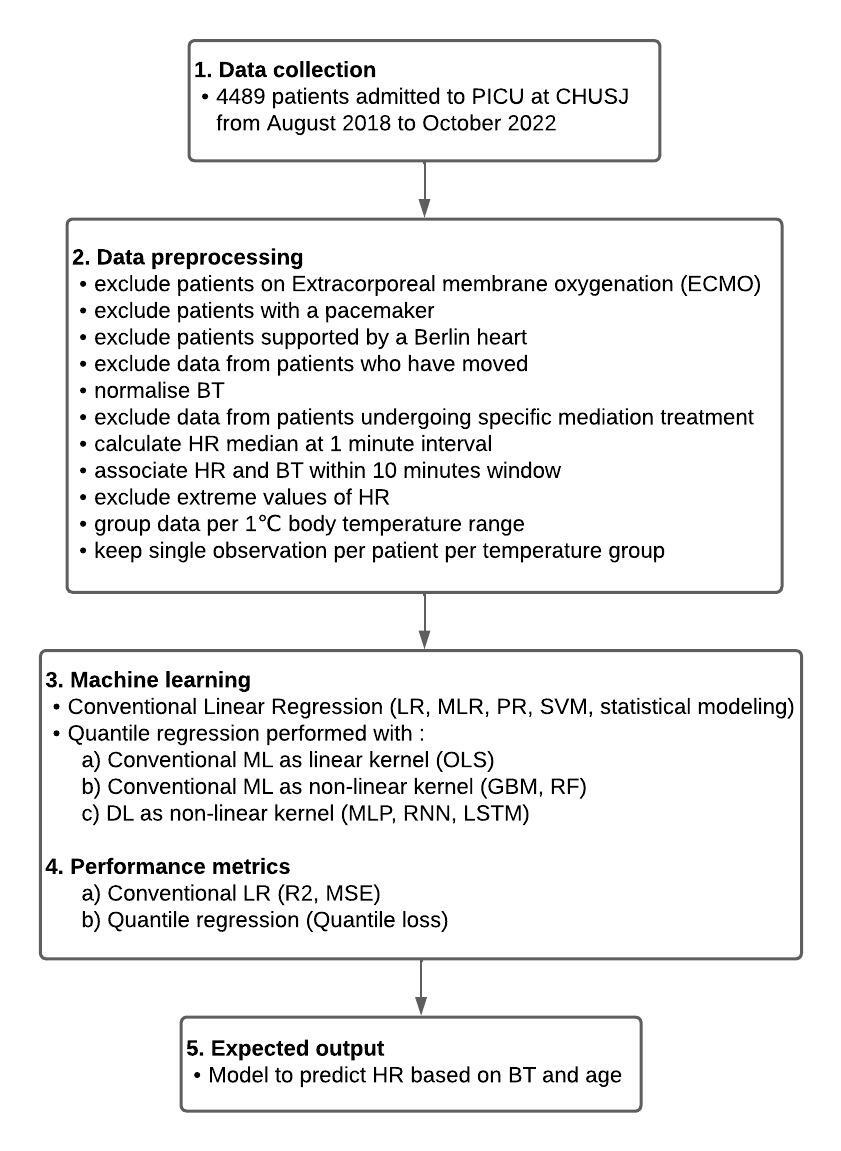}
    \caption{Workflow to model the relationship between HR and BT.}
    \label{fig:flowchart}
\end{figure}

\subsection{Data collection}
Since 2015, a high-resolution database has been in operation at CHUSJ, collecting clinical data from patients admitted to the PICU at 1-second to 5-second intervals from monitors. Patients were continuously monitored during their stay until their discharge \cite{brossier2018creating}. This project aims to establish a relationship between HR and BT. Patients at CHUSJ PICU were closely monitored for both physiological markers. For BT, continuous monitoring occurred at 30-second intervals, complemented by manual measurements taken every 2-4 hours. The measurement method varied based on the patient's age and included using rectally, esophageal, or axillary temperature probes. HR was monitored using electrodes on the patient's chest, generating an electrocardiogram (ECG). Alternatively, HR could also be obtained using a pulse oximeter attached to the patient's finger, which provides oxygen saturation data and HR measurements.

The eligible population included all children aged 0 to 18 admitted to the PICU between August 2018 and October 2022, for whom HR and BT records were available. Certain patient groups were excluded from the study in this population to avoid potential confounding factors that interact with BT and HR values. Therefore, removing these data points was essential to achieve accurate modeling. Patients on extracorporeal membrane oxygenation (ECMO), with a pacemaker, or supported by a Berlin heart were excluded. By excluding those patients, our data extraction process yielded information from 4,007 patients admitted between August 2018 and October 2022. From this patient cohort, we extracted patient age at admission, 4 days (96 hours) of HR and BT (value, temperature site, measurement type such as continuous or manual) data along with the corresponding date and time of acquisition, comfort scores and specific medication treatment with the day when the drug treatment was finished. 

\subsection{Data preprocessing}
Data preprocessing is a fundamental initial step in preparing patient data for the ML algorithm. Effective data preprocessing ensures data quality issues that can impact the validity of any conclusions drawn \cite{kinaneva2021machine}. To get better reliability of subsequent insights, the process involves the following tasks: 

\subsubsection{Exclude data from patients who have moved} Child's movements, including when a patient is restless, crying, or screaming, were excluded from the study. The analysis of the following comfort scores assisted in identifying and excluding data from patients who have moved : 
\begin{enumerate}
    \item \textbf{Cornell Assessment of Pediatric Delirium (CAPD)} scale was used in PICU for invasively ventilated children who have fluctuations in awareness, attention, and cognition \cite{cloedt2021impact}. The evaluation had 8 questions, and each answer was associated with a score between 0 and 4. The total of these scores determined the outcome; delirium would be absent if the scores were less than 9. 
    \item \textbf{Comfort Behavior (Comfort B)} was used to assess pain in ventilated and sedated patients \cite{amigoni2022recommendations}. The score involved observing 6 behaviors (awakening, agitation, ventilation, movement, facial expression, muscle tone) rated on a scale of 1 to 5 to determine the perceived pain level. A total score between 11 to 17 indicated a normal score where the patient had no pain.  
    \item \textbf{FLACC (Face, Legs, Activity, Cry, Consolability)} scale was used to evaluate pain in noninvasive patients under 6 years \cite{cloedt2021impact}. Each behavior was scored on a scale from 0 to 2, and patients were included in our study if they had a total score between 0 and 3. This range indicated the absence of pain or mild pain.
    \item \textbf{revised FLACC (r-FLACC)} was used for patients with intellectual disability \cite{cloedt2021impact}. The evaluation was executed the same way as the FLACC scale. 
    \item \textbf{Visual numeric Scale (VNS)} was used to evaluate and monitor pain for patients over 6 years with the ability to communicate \cite{amigoni2022recommendations,cloedt2021impact}. The patient evaluated the pain on a scale of 10, with a score between 0 and 3 indicating the absence of pain or mild pain.
    \item \textbf{Richmond Agitation-Sedation Scale (RASS)} was a scale used to evaluate agitation and sedation \cite{cloedt2021impact}. We only kept data with scores between -5 and +1, which signified instances of physical and verbal stimulation. 
\end{enumerate}

Based on scores extracted from the database, we assigned movement or not based on the score. Afterward, we associated the scores with the nearest date-time values of BT value since it was monitored at a lower frequency (hours) than HR (seconds). If any of the scores indicated the presence of movement, we excluded the data associated with those movement instances to work with data from calm and non-moving patients. 

\subsubsection{Normalize body temperature}
We normalized the temperature data, specifically for measurements obtained from the axillary site, to ensure consistency and accuracy in our analysis.  The interpretation of axillary temperatures was underestimated by approximately 0.5°C compared to other methods. Temperature documentation in the PICU was done by recording the actual temperature value displayed on the thermometer. The site from which the temperature was taken (oral, axillary, or rectal) was also noted. Among these methods, rectal temperature measurement was usually considered the most precise \cite{dolibog2022comparative}.

\subsubsection{Exclude data from patients undergoing specific medication treatment}
Patients in the PICU who received Dexmedetomidine and medication affecting HR were only included once their medication treatment was completed. Specifically, medications used to treat heart conditions that slow down HR included antiarrhythmics, beta-blockers, calcium channel blockers, digoxin, and ivabradine \cite{meyer2018pharmacological,king2018antiarrhythmic}. Conversely, medications that increased HR included Dobutamine, Dopamine, Epinephrine, Milrinone, Norepinephrine (or Noradrenaline), and Salbutamol. To exclude the medication administration period, we excluded data when the monitored timestamp of vital signs coincided with the medication administration.

\subsubsection{Calculate heart rate median at 1-minute interval}
In the PICU, HR values were recorded initially every second for all patients. Given the high frequency of measurements, we simplified the data volume by doing data aggregation. More specifically, calculating the median HR at one-minute intervals made the original high-frequency HR data more manageable and insightful.

\subsubsection{Associate HR and BT within 10 minutes window}
Establishing the association between HR and BT was important as it provided valuable insights into the patient's physiological response. Since HR and BT were monitored at different PICU intervals (seconds vs. hours), we associated these two variables by calculating the median of HR measurements within a 10-minute window, more precisely within $\pm$ 5 minutes of each BT timestamp.

\subsubsection{Exclude extreme values of heart rate}
Extreme HR measurements—values lower than 30 or higher than 240 bpm—were excluded from the dataset. We hypothesized that these numbers could result from unusual clinical diseases or data entry errors, which could have introduced inaccuracies in the analysis. By removing these outliers, we wanted to ensure our dataset was more resilient and reliable. This would have allowed us to base future analysis and modeling efforts on a more accurate depiction of physiological parameters within the target population.

\subsubsection{Group data per 1°C body temperature range}
We segmented the data into temperature ranges of 1 degree Celsius
(°C) each to conduct specific temperature value analyses. This
temperature-based segmentation created groups ranging from
33°C to 33.9°C up to 40°C to 40.9°C. By doing so, we
performed detailed investigations into how patient's physiological
responses vary across different temperature ranges, providing
insights that could be valuable for clinical and research purposes.

\subsubsection{Keep single observation per patient}
To ensure the integrity and balance of our dataset, we addressed the initial variability in the number of data points attributed to each patient. Initially, there was a substantial variance in the volume of data for each patient. To rectify this imbalance, we implemented a data reduction strategy. We preserved only 1 observation for each patient per temperature group to maintain meaningful and representative data. This practice simplified our dataset and ensured each patient's contribution to the analysis was meaningful and equitable.

After preprocessing, our dataset comprised 4462 pairs of HR-BT values represented in Table \ref{table:observations}. The values were organized based on BT values, where we used temperature grouping with 1-degree increments. This decision allows for a more granular representation of temperature variations and makes it easier for the ML algorithm to identify patterns in the data \cite{kinaneva2021machine}. Moreover, the table provides a breakdown of patient counts according to the patient's age, offering additional insights into the age distribution within the dataset. The age groups include newborn (0-28 days), infant (29 days - 1 year), toddler (1-2 years), child (2-12 years) and teenager (12-18 years). This meticulous grouping strategy corresponds to the Food and Drug Administration (FDA) standard age categorization, with a more detailed grouping specifically for patients aged 1 to 2 years \cite{fda_pediatric_medical_devices}. It also facilitates a more nuanced analysis, considering temperature variations and age-specific patterns in pediatric patients.

\begin{table}[h!]
    \centering
    \scriptsize  
    \caption{Number of observations for each body temperature category by age group following data preprocessing}
    \label{table:observations}
    \setlength{\tabcolsep}{3pt}  
    \begin{tabular}{||c c c c c c c||} 
        \hline
        BT range (°C) & Newborn & Infant & Toddler & Child & Teenager & Total\\ [0.5ex] 
        \hline\hline
        33-33.9 & 2 & 2 & 0 & 1 & 1 & 6\\ 
        34-34.9 & 8 & 3 & 0 & 4 & 2 & 17\\
        35-35.9 & 10 & 15 & 3 & 18 & 13 & 59\\
        36-36.9 & 110 & 335 & 139 & 640 & 391 & 1615\\
        37-37.9 & 136 & 442 & 201 & 752 & 420 & 1951\\
        38-38.9 & 48 & 168 & 53 & 243 & 117 & 629\\
        39-39.9 & 5 & 28 & 13 & 78 & 28 & 152\\
        40-40.9 & 0 & 3 & 2 & 21 & 7 & 33\\
        [0.5ex] 
        \hline
    \end{tabular}
\end{table}

\subsection{Machine Learning}

Following the preprocessing steps, the objective is to capture the intricate relationship between HR and BT in critically ill pediatric patients. We will validate this finding with our data by employing conventional linear regression models based on previous studies highlighting linear relationships. For this study, four popular algorithms were used for modeling the linear relationship \cite{kinaneva2021machine}: 

\subsubsection{Conventional Linear Regression} 
\begin{enumerate}[(a)]
    \item \textbf{Linear Regression (LR)} is used to determine the best linear function that fits a given set of input-output pairs \cite{kinaneva2021machine}. This approach was also used by Kirschen, Singer, Thode Jr, \& Singer, 2020 and Jensen \& Brabrand, 2015 where they respectively observed that a 1°C rise in BT would cause a 7 bpm and a $7.4 \pm 0.9$ bpm increase in HR among adults admitted to the emergency department \cite{kirschen2020relationship,jensen2015relationship}. The limitation of this technique is only one independent variable can be used to predict the dependent variable, making this approach less accurate for prediction. \cite{kinaneva2021machine}. 
    \item \textbf{Multiple Linear Regression (MLR)} is also used to build a linear model to predict the dependent variable related to one or many independent variables, which makes this technique a better prediction model than the linear regression \cite{kirschen2020relationship}. This technique was used by Heal et al., 2022, Jensen \& Brabrand, 2015 and Broman, Vincent, Ronco, Hansson, \& Bell, 2021 where they all concluded in a linear change between HR, BT and age \cite{heal2022association,jensen2015relationship,broman2021relationship}. 
    \item \textbf{Polynomial Regression (PR)} with first degree, is a linear regression equation. The first-degree polynomial, often called simple LR, considers terms up to the first power, resulting in a model where the response variable is a linear function of the predictors. \cite{kinaneva2021machine}
    \item \textbf{Support Vector Machine (SVM)} can be used for both classification and regression problems. In this work, we employed the linear kernel, and the objective was to fit as many data points as possible to the hyperplane line inside the demarcation lines, known as the decision boundaries. This approach's strength is its excellent prediction accuracy and robustness against outliers. However,  it does not attempt to minimize the errors between the true and predicted values, in contrast to other regression algorithms \cite{kinaneva2021machine}. 
    \item \textbf{Statistical modeling} approach by Davies \& Maconochie, 2009 enabled them to predict HR at different quantiles, leading to the conclusion that "BT serves as an independent determinant of HR, resulting in an approximate increase of 10 bpm per degree centigrade" \cite{davies2009relationship}. In our study, we used their established equation to determine its coefficients, then evaluated its effectiveness in predicting HR with our data \cite{davies2009relationship}. Their statistical modeling is shown below : 

\begin{small}
\begin{align}
    \text{Expected HR value} = \text{BT} \cdot a + \text{Age} \cdot b + (\text{Age}^2) \cdot c + \text{constant}
\end{align}
\end{small}

    where $BT$ is the body temperature in °C and $Age$ is the age in month \cite{davies2009relationship}.

\end{enumerate}

\subsubsection{Quantile regression}:
Furthermore, we incorporated Quantile regression (QR) into our analysis for a more comprehensive exploration, combined with various traditional ML algorithms and DL models. This method allowed us to go beyond linear assumptions and enhanced our ability to uncover linear and non-linear patterns in our dataset. We also included interaction terms BT and age to enhance the accuracy of the predictive model for HR and provide a more comprehensive understanding of the dynamic.

QR is a statistical method used to estimate the response variable at various quantiles (e.g., 0.10, 0.25, 0.50, 0.75, 0.90). Its primary purpose is to analyze the impact of predictor variables (independent variable) on the output variable (dependent variable) across these quantiles. Unlike traditional regression methods, such as Least squares regression, which focus on estimating the mean, QR can offer insights into how the predictors influence various parts of the response distribution, making it valuable in situations where the relationship is not uniform across all quantiles and suitable for capturing a broader range of information about the response variable \cite{patidar2023quantile}. This method is typically utilized when the assumptions of linear regression are unmet. One of the advantages of QR lies in its robustness, as it is less sensitive to outliers compared to the least squares method. This robustness enhances the reliability of the technique in the presence of atypical data points. Additionally, QR offers flexibility by enabling the capture of the entire distribution of the outcome variables, providing a more comprehensive view of the relationship between variables. Moreover, it addresses the issues related to heteroscedasticity, where the variance of the dependent variable varies differently depending on the levels of the independent variable. This implies that the variance of residuals is not constant across all levels of the response variable \cite{patidar2023quantile}. Another benefit of using QR is that we may customize the kernel, which allows us to improve our capacity to identify and comprehend subtleties in the relationships between variables. Because of its adaptability, we continue to analyze three different kernels: conventional ML as a linear kernel, conventional ML as a non-linear kernel, and DL as a non-linear kernel.

\begin{enumerate}[(a)]
    \item {Conventional ML as a linear kernel}
    \begin{enumerate}[(i)]
        \item \textbf{Ordinary Least Squares (OLS)} is utilized to model and estimate the coefficients of a linear relationship between variables, particularly when the assumed relationship is linear. OLS aims to comprehend the correlation between variables and make predictions while minimizing the sum of all squared, also called the Residual sum of squares (RSS), differences between observed and predicted values. This method proves efficient in providing accurate parameter estimates when the assumption of linearity holds. It also offers interpretability where the coefficients present clear insights into the relationship between the variables \cite{hutcheson2011sage}.
    \end{enumerate}

    \item {Conventional ML as a non-linear kernel}
    \begin{enumerate}[(i)]
        \item \textbf{Gradient Boosting Machine (GBM)} creates a predictive model in the form of a decision tree, where each tree in the ensemble corrects the errors of its predecessor, leading to an overall model with enhanced accuracy and robustness. The primary objective of GBM is to generate a predictive model capable of capturing complex relationships in the data. This versatile technique can be applied to both regression and classification problems. One of its key strengths lies in its iterative improvement process, where subsequent trees focus on addressing the residuals left by previous ones. Its iterative and sequential construction allows it to capture and model complex, non-linear interactions between variables. This flexibility makes GBM a valuable tool for tasks where the underlying relationships may not follow a simple linear structure. Some notable advantages of GBM include high predictive accuracy, the capability to handle both linear and non-linear relationships effectively, and robustness against overfitting. The model's ability to learn from its mistakes and continuously refine its predictions makes it a valuable tool for tasks requiring precision and adaptability in the face of complex data structures \cite{fafalios2020gradient,aziz2020study,boehmke2019hands}.
        \item \textbf{Random Forest (RF)} constructs many decision trees, each built on a subset of the data during training. Consequently, the output represents these trees' mean or average prediction for regression problems. Each tree focuses on a specific subset, and the final prediction aggregates these individual tree predictions. The algorithm offers several benefits: flexibility to handle regression and classification problems, robustness against overfitting compared to decision trees, ability to highlight feature importance for better predictive performance, effectiveness in capturing complex relationships and non-linear patterns, and reduced sensitivity to outliers \cite{cutler2012random}.
    \end{enumerate}

    \item {DL as a non-linear kernel} 
    \begin{enumerate}[(i)]
        \item \textbf{Recurrent Neural Network (RNN)} is designed to discover patterns in sequential data, making them suitable for time series analysis and natural language processing. Their fundamental advantage lies in their ability to use their collection of previous inputs to make real-time decisions by doing the same task on each sequence element. This model's incapacity to provide background knowledge over an extended period of time is one of its limitations \cite{sivamohan2021effective}.
        \item \textbf{Long Short-Term Memory (LSTM)}, an advanced form of RNN, addresses the issue of learning long-term dependencies. Its unique architecture incorporates four interacting layers containing chains of repeated modules. This model's memory cells enable it to store and retrieve information selectively over a longer time. This makes LSTMs highly effective in tasks requiring understanding context over extended periods \cite{sivamohan2021effective, saha2020long}.
        \item \textbf{Multilayer Perceptron (MLP)} is characterized by three layers (input, hidden, and output layers) of interconnected nodes (neurons), where each node in one layer is linked to every node in the next layer with associated weights. Also known as a supplement of feed forwards NN, the data move in one direction from the input, hidden to the output layer. The multiple hidden layers represent the computational engine of MLP. It facilitates feature learning, allowing the network to acquire hierarchical representations of data and capture intricate patterns and features due to their nonlinear characteristics. Moreover, The MLP's adaptability and effectiveness make it a popular choice for various ML applications, including regression and classification \cite{molina2021classification}.
    \end{enumerate}
\end{enumerate}

\subsection{Performance metrics}

The following performance metrics were used to assess the effectiveness of ML and DL models. We used R-squared and Mean square errors to evaluate conventional linear regression. While QR models were evaluated using quantile loss. These metrics provided a comprehensive evaluation of the models' performance, offering insights into their ability to predict and capture the underlying patterns within the data. Each metric contributed to a nuanced understanding of the model's accuracy for the regression task \cite{varoquaux2023evaluating}. 

\textbf{- R-squared (R2)}, or the coefficient of determination, represents "the proportion of variance in the dependent variable (output) that is predictable from the independent variables (input)" \cite{chicco2021coefficient}. The interpretation of R2 provides a quantitative measure of the quality fit. Potentially ranging from 0 to 1, an R2 value near 1 indicates that the model explains all of the dependent variable's variability, signifying a robust fit. In contrast, an R2 value nearing 0 indicates that the model fails to explain the variability in the dependent variable, implying a less effective fit \cite{keer2023goodness}. The R2 equation is defined below:

\begin{align}
    R^2 = 1 - \frac{\sum_{i=1}^n (Y_i - \hat{Y_i})^2}{\sum_{i=1}^n (Y_i - \tilde{Y})^2}
\end{align}

where $n$ is the number of observations, $Y_i$ is the true value, $\hat{Y_i}$ is the predicted value and $\tilde{Y}$ is the mean of $Y$ values. More precisely, the numerator is the sum of the squared residual (SSR), representing the difference between the actual and predicted values. The denominator is the total variations of the sum of squares, representing the sum of the distance between the data and the mean all squared \cite{keer2023goodness}.

\textbf{- Mean Square Error (MSE)} measures the average squared difference between the true and the predicted values. Being only a positive value, the interpretation of the MSE value is as follows: a lower MSE signifies that the predicted values exhibit closer proximity to the actual values, reflecting an enhanced overall fit. This metric provides a standardized measure for evaluating regression models' precision and accuracy, facilitating straightforward comparisons between different models \cite{tyagi2022regression,rane2023optimal}. The MSE equation is defined below : 

\begin{align}
    MSE = \frac{1}{n} \sum_{i=1}^{n}(Y_i - \hat{Y_i})^2
\end{align}

where $n$ is the number of observations, $Y_i$ is the true value and $\hat{Y_i}$ is the predicted value. \cite{tyagi2022regression}

\textbf{- Quantile Loss (QL)}, also known as Pinball Loss, is a metric used in QR to calculate the performance of conditional quantiles, providing insight into how well a model captures the distribution of the data \cite{steinwart2011estimating}. Unlike other ML algorithms, R2 is not a good metric to evaluate performance since it is not based on distribution assumptions \cite{machado2013quantile}. A reminder that QR is a statistical method that aims to model the response variable's conditional distribution through different quantiles, allowing us to have information on the entire distribution, not just the mean. To interpret QL results, a low QL value means a superior performance from the model, suggesting that the predicted values closely align with the actual values at the designated quantile \cite{narayanexpected}. The QL equation is shown below :

\begin{align}
    QL_\tau (y,\hat{y}) = max(\tau (y - \hat{y}),(\tau - 1)(y - \hat{y}) )
\end{align}

where $\tau$ is the quantile level, $y$ is the true value, $\hat{y}$ is the predicted value \cite{narayanexpected}. To compare the performance of quantile regression for each model, it is necessary to compute the total quantile loss, achieved by calculating the average quantile loss across all quantiles. This enables the performance analysis of all quantiles within a model \cite{alutaibi2020network}.

In summary, the ultimate goal is to identify patterns and associations and develop a robust predictive model. The model aims to assist healthcare professionals in anticipating changes in HR based on variations in BT, age, and quantile of interest. Through the integration of advanced ML methodologies, this study seeks to propel the field of pediatric critical care forward, providing clinicians with actionable insights to tailor interventions and enhance overall patient care in the PICU.

\section{Results and discussion}

\begin{figure*}[h]
\centerline{\includegraphics[scale=0.58]{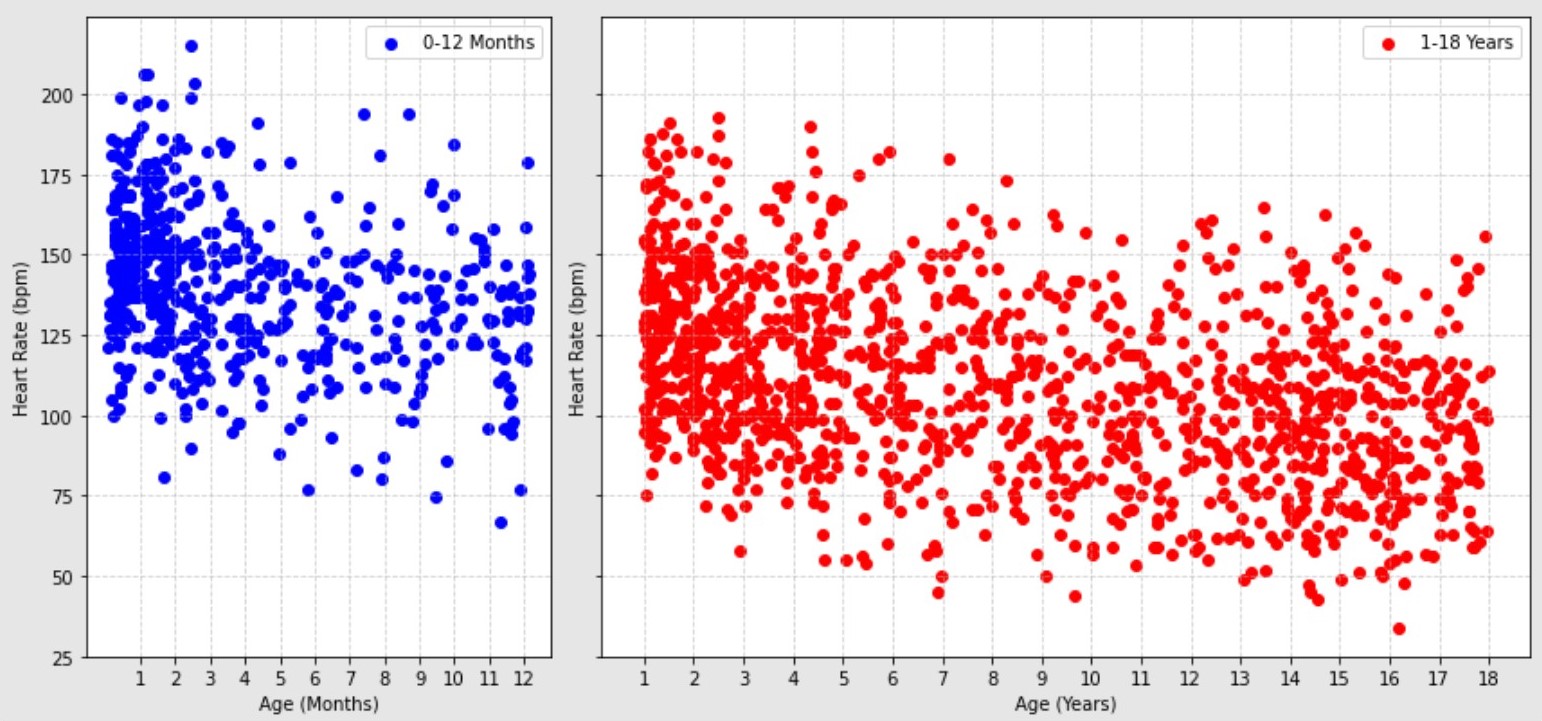}}
\caption{Heart rate as a function of age for children from 0-18 years old admitted to the PICU with body temperature between 37 to 37.9°C}
\label{fig_hr_age_37}
\end{figure*}

Firstly, our initial findings from the research involve the generation of a graph showing HR concerning age for a temperature range presented in Figure \ref{fig_hr_age_37}. Upon analysis, we observe a trend where HR decreases with the patient's age. This observation aligns with the findings presented in the article by the study \cite{fleming2011normal}. The graphical representation provides valuable insights into the relationship between age and HR within distinct temperature ranges, highlighting a potential correlation that needs further investigation. This initial exploration sets the foundation for a more in-depth analysis of age-related variations in HR across different BT intervals.

Expanding on these insights, we applied ML techniques to uncover subtle patterns and dependencies within our dataset. Each algorithm underwent meticulous hyperparameter tuning to optimize the model performance. Hyperparameters were set before the training process. Meticulous tuning suggests a careful and detailed examination of these hyperparameters to achieve the best possible performance from the model. It involves iteratively adjusting individual hyperparameters, training the model, and evaluating its performance until an optimal configuration is found. The goal is to enhance the model's effectiveness and accuracy. This meticulous approach to hyperparameter tuning is essential for achieving optimal results and maximizing the model's potential.

Since prior research has consistently underscored the linear association between HR and BT \cite{daymont2015heart,davies2009relationship,thompson2009deriving,heal2022association,kirschen2020relationship,jensen2015relationship,broman2021relationship}, we initially used conventional linear regression models. We started with LR to predict HR based on age for each BT range. Subsequently, we explored MLR and PR degree 1 to incorporate BT and age as independent variables. Finally, we experimented with SVM with a linear kernel and the statistical model. Examining the outcomes of our analyses, the performances of these traditional linear regression methods in capturing the linear relationship for BT ranging from 37 to 37.9°C are summarized in Table \ref{table:model_performance_linear}. From the comparative analysis, we can conclude that there are no clear and obvious linear relations between HR and BT because of significantly small R2 values and a large value of MSE loss. These values suggest that the linear models struggled to account for a substantial portion of the variability in HR based on BT and age. Based on these findings, there may be a potential need for more advanced methodologies to effectively capture the intricacies of the relationship between HR and BT.

Our experimental findings suggest that linear models might not be able to adequately capture the complexity of the relationship observed in critically ill patients. This assertion is supported by the studies \cite{momo2023measuring, heal2023response}, highlighting the non-linear correlation between HR and BT. Specifically, in Momo et al. (2023) article \cite{momo2023measuring}, the authors critique the work of Heal et al. (2022)  \cite{heal2022association}, who utilized linear regression models to demonstrate a significant increase in HR of 12.3 bpm for every 1°C rise in BT among emergency department patients. Additionally, Momo et al. (2023) underscore the significance of findings from other studies employing non-linear models. These studies reveal a distinct, curve-like increase in HR within the temperature range of 37°C to 38°C, highlighting the intricate relationship between HR and BT \cite{momo2023measuring}.

\begin{table}[h!]
\caption{Model performance from traditional linear Machine Learning techniques to predict heart rate from children 0-18 years old with body temperature between 33 to 40.9°C}
\label{table:model_performance_linear}
\centering
\begin{tabular}{||c c c||} 
 \hline
 Conventional linear regression & R2 & MSE\\ [0.5ex] 
 \hline\hline
 Linear regression & 0.3145 & 621.1870\\ 
 Multiple Linear Regression & 0.3563 & 583.2989\\
 Polynomial regression (degree = 1) & 0.3563 & 583.2989\\
 SVM (kernel: linear) & 0.3152 & 620.5312\\
 Statistical model (quantile=0.5) & 0.3576 & 581.2691 \\
[0.5ex] 
 \hline
\end{tabular}
\end{table}

\begin{figure*}[h!]
\centerline{\includegraphics[scale=0.5]{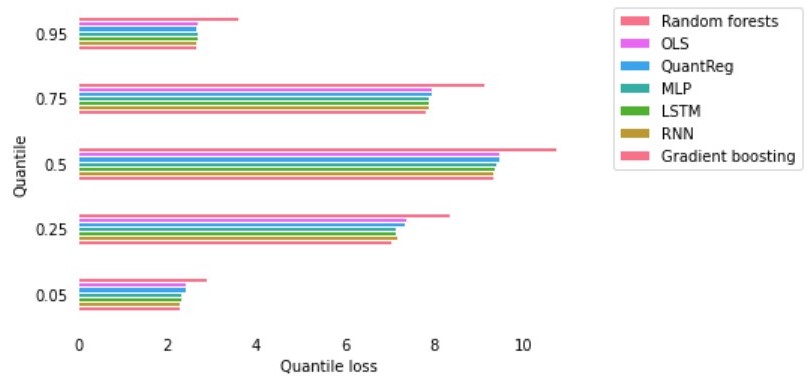}}
\caption{Comparison of quantile loss per quantile for each technique for a body temperature range between 33 to 40.9°C from critically Ill patients 0-18 years old from 1 experiment}
\label{fig_quantile_loss_quantile}
\end{figure*}

The authors respond to the previously mentioned critiques in the article by Heal et al. (2023). They acknowledge that employing linear models may not be the most accurate approach and suggest using polynomial regression. They also note that contrary to earlier research suggesting a general 1°C increase correlating with a 10 bpm rise, their findings indicate a 12.3 bpm increase, emphasizing potential variations within specific temperature ranges. This distinction gains particular significance when considering the application of such predictive tools to individual patients in a clinical setting \cite{heal2023response}.

\begin{figure}[h!]
	\centering
	
	\begin{minipage}[t]{0.295\textwidth}
		\centering
		\includegraphics[width=\textwidth]{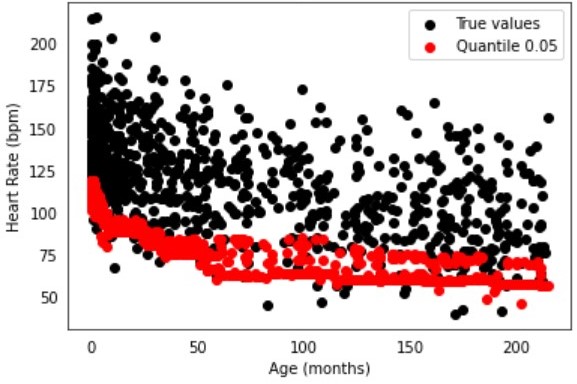}
		\caption*{(A)}
	\end{minipage}

	\hfill
	
	\begin{minipage}[t]{0.295\textwidth}
		\centering
		\includegraphics[width=\textwidth]{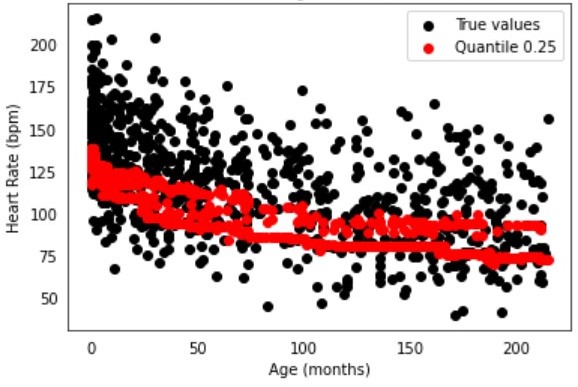}
		\caption*{(B)}
	\end{minipage}

	\vspace{0.01em}  
	
	\begin{minipage}[t]{0.295\textwidth}
		\centering
		\includegraphics[width=\textwidth]{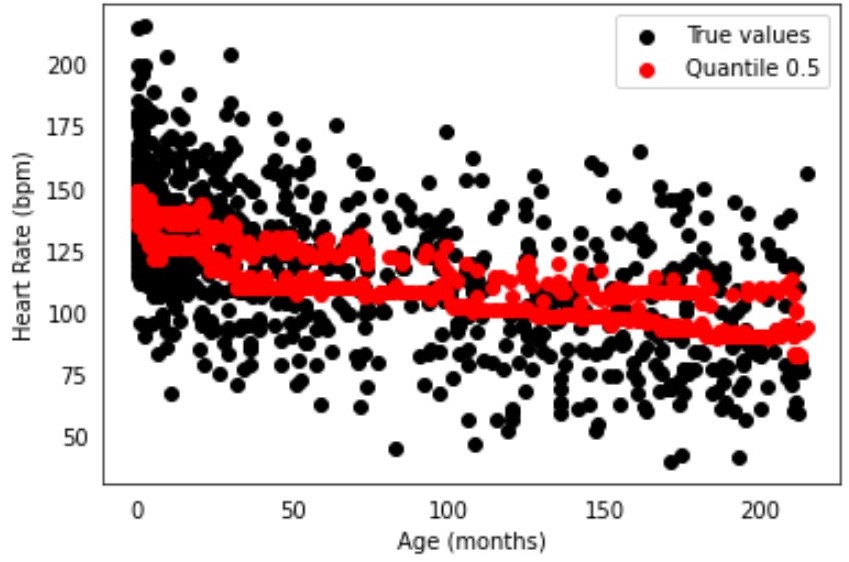}
		\caption*{(C)}
	\end{minipage}

	\vspace{0.01em}  
	
	\begin{minipage}[t]{0.295\textwidth}
		\centering
		\includegraphics[width=\textwidth]{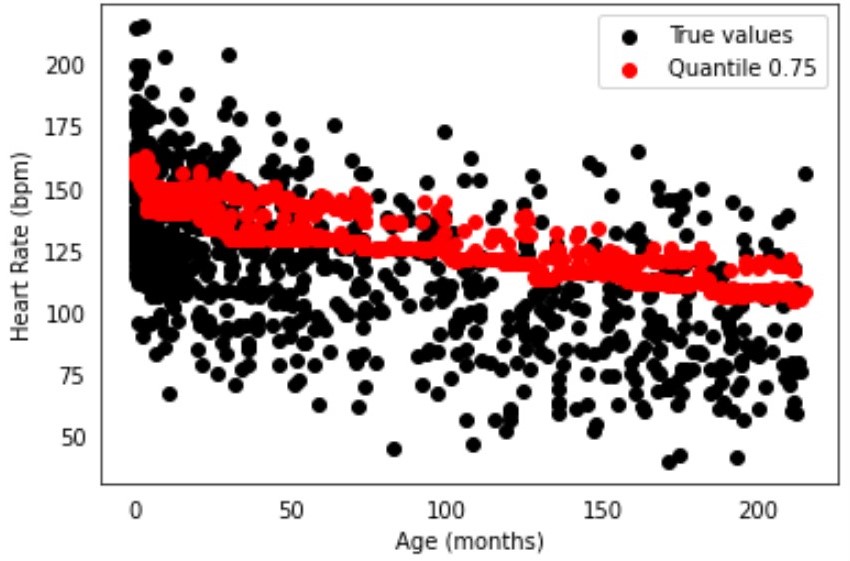}
		\caption*{(D)}
	\end{minipage}

	\vspace{0.01em}  

	\begin{minipage}[t]{0.295\textwidth}
		\centering
		\includegraphics[width=\textwidth]{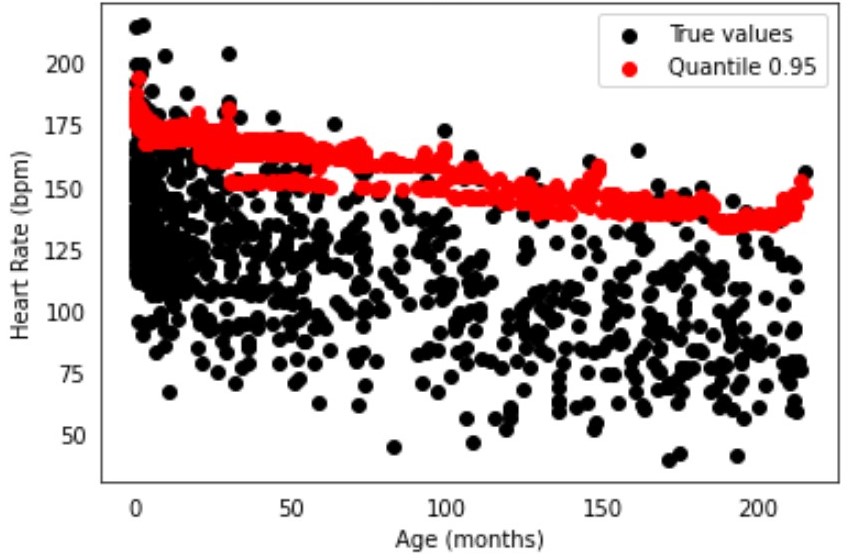}
		\caption*{(E)}
	\end{minipage}

	\caption{QR performed with GBM model shows a non-linear relationship between HR, BT, and age for children 0-18 years old admitted to PICU with a temperature range of 34 to 40.9°C. True values are shown in black, while predictions are shown in red. The analysis is performed at different quantiles, where each row represents a specific quantile, from 0.05 to 0.95. Subfigure (A) illustrates model predictions at quantile 0.05, (B) at quantile 0.25, (C) at quantile 0.50, (D) at quantile 0.75, and (E) at quantile 0.95. }
	\label{fig:qr_gbm_non_linear_all_quantiles_temp_33_40}
\end{figure}

\begin{figure}[h!]
	\centering
	
	\begin{minipage}[t]{0.29\textwidth}
		\centering
		\includegraphics[width=\textwidth]{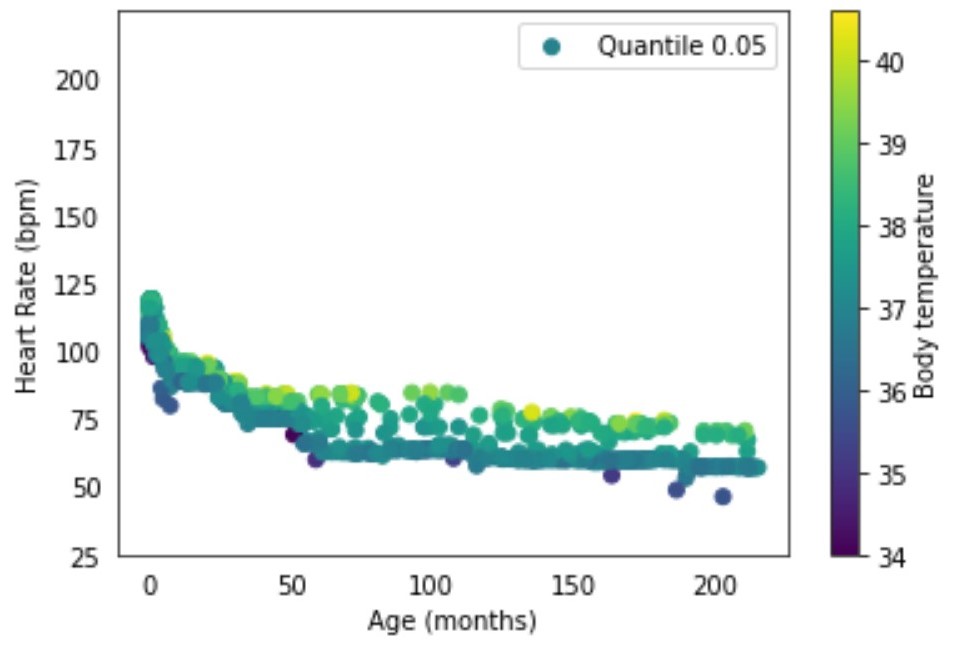}
		\caption*{(A)}
	\end{minipage}

	\hfill
	
	\begin{minipage}[t]{0.29\textwidth}
		\centering
		\includegraphics[width=\textwidth]{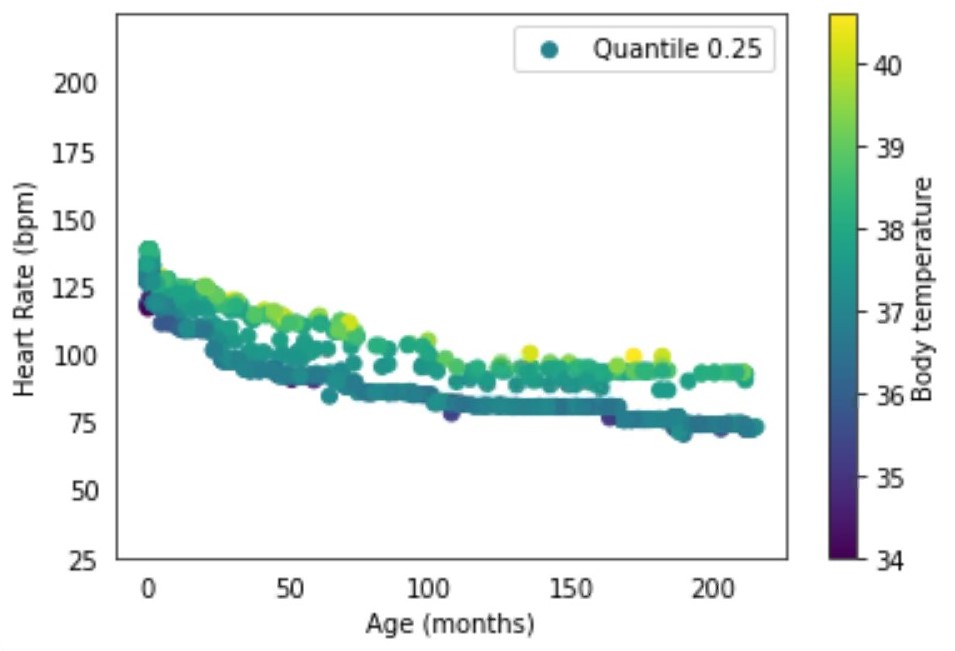}
		\caption*{(B)}
	\end{minipage}

	\vspace{0.01em}  
	
	\begin{minipage}[t]{0.29\textwidth}
		\centering
		\includegraphics[width=\textwidth]{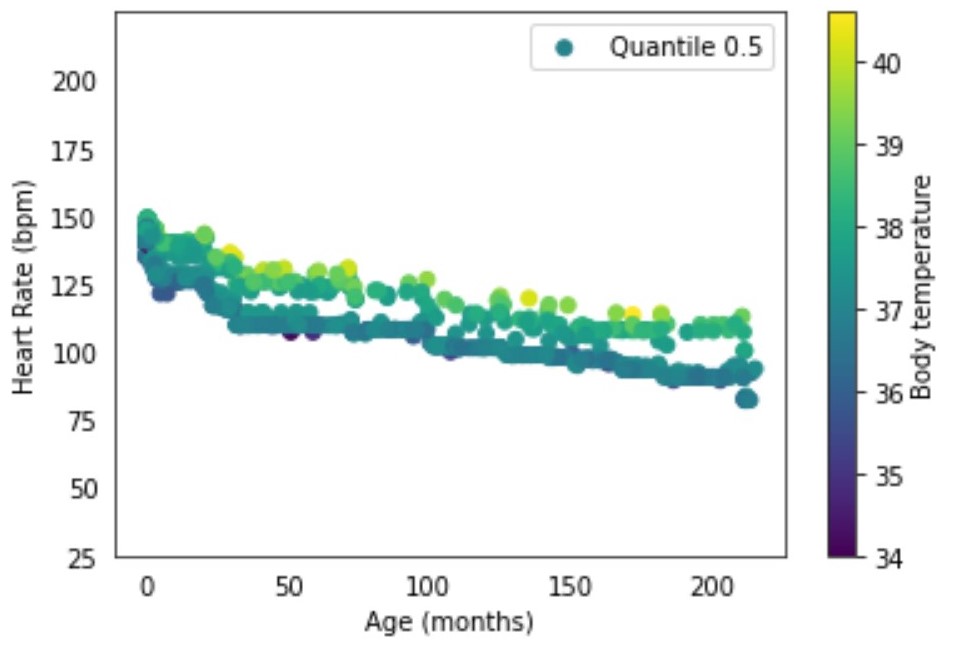}
		\caption*{(C)}
	\end{minipage}

	\vspace{0.01em}  
	
	\begin{minipage}[t]{0.29\textwidth}
		\centering
		\includegraphics[width=\textwidth]{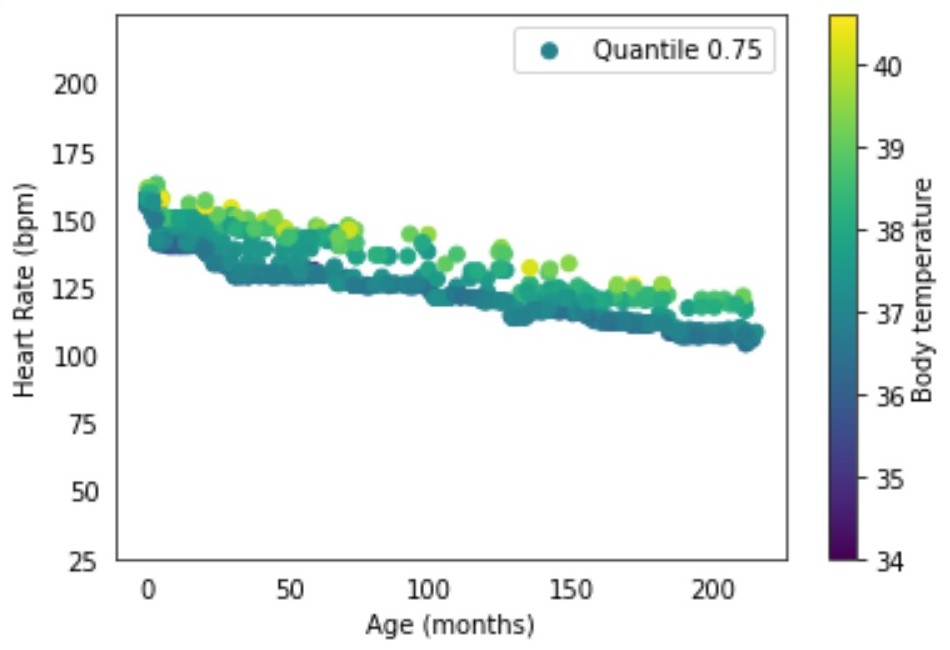}
		\caption*{(D)}
	\end{minipage}

	\vspace{0.01em}  

	\begin{minipage}[t]{0.29\textwidth}
		\centering
		\includegraphics[width=\textwidth]{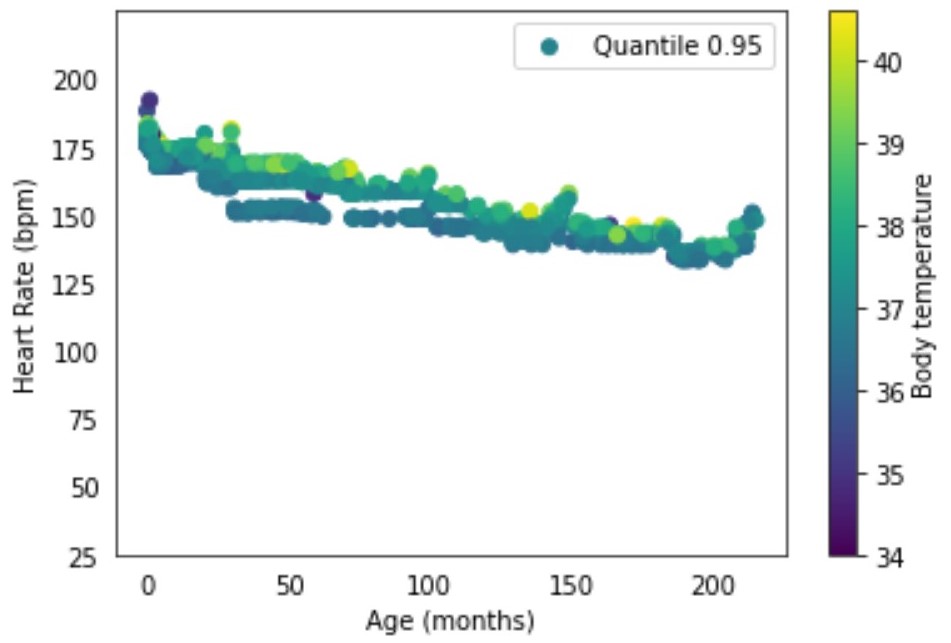}
		\caption*{(E)}
	\end{minipage}
	
	\caption{Exploration of the HR model predictions influenced by age along the x-axis and BT through variation of 34 to 40.9°C in the color scale for children 0-18 years old admitted to PICU. The analysis is performed at different quantiles, where each row represents a specific quantile, from 0.05 to 0.95. Subfigure (A) illustrates model predictions at quantile 0.05, (B) at quantile 0.25, (C) at quantile 0.50, (D) at quantile 0.75, and (E) at quantile 0.95.}
	\label{fig:model_predictions}
\end{figure}

Since traditional linear models do not adequately capture the complex and potentially non-linear relationship, we conducted a comprehensive evaluation using QR performed with ML and DL as a linear and non-linear kernel. Through this, we aim to enhance the depth and precision of our analysis. The advantage of QR lies in its ability to handle data with varying distributions and to provide a more nuanced understanding of the conditional distribution of the response variable across different quantiles \cite{patidar2023quantile}. This is particularly relevant when the relationship between variables may not be constant across the entire distribution. To evaluate the models with QR, it is unsuitable to use R2, as mentioned in the literature \cite{machado2013quantile}. Instead, it is essential to investigate the quantile loss (QL) for each quantile as the performance metric for QR. The results presented in Figure \ref{fig_quantile_loss_quantile} reveal that the lowest QL value is seen at each quantile in QR performed with GBM, showing that the underlying relationship is effectively captured.

To assess the performance of each model with QR, the performance metric is the total quantile loss, representing the overall performance across all quantiles within a model. This metric is obtained by calculating the average QL across all quantiles. The data presented in Table \ref{table:total_quantile_loss} detail the average total quantile loss and the standard deviation (SD) for each predictive model derived from the outcomes of five distinct experiments. The best model is once again QR performed with GBM. The model demonstrates superior efficacy by achieving the lowest total quantile loss value. The superior performance of this model suggests its efficacy in handling the complexities in the data, contributing to enhanced predictive capabilities. This advanced method outperforms conventional approaches with linear kernels, such as OLS, RF, and standalone QR. Specifically, the higher total quantile loss result for RF implies potential limitations in its effectiveness within this context. These results underscore the significance of employing advanced ML and DL techniques for capturing nuanced patterns in critical care scenarios, providing valuable insights into the relationships between physiological parameters.

\begin{table}[h!]
\centering
\caption{Mean total quantile loss and SD from 5 experiments for different methods performed with QR with a temperature range between 33 to 40.9°C from critically Ill patients 0-18 years old}
\label{table:total_quantile_loss}
\begin{tabular}{||c c||} 
 \hline
 Model & Mean total quantile loss $\pm$ SD \\ [0.5 Ex] 
 \hline\hline
GBM & 6.5069 $\pm$ 5.2507e-05\\
RNN & 6.5633 $\pm$  0.0078\\
LSTM & 6.5990 $\pm$ 0.0144 \\
MLP & 6.6148 $\pm$ 0.0236 \\
OLS & 6.6867 $\pm$ 0.0 \\
QR & 6.6903 $\pm$ 0.0 \\
RF & 7.6584 $\pm$ 0.0 \\ [0.5ex] 
 \hline
\end{tabular}
\end{table}

Fig. \ref{fig:qr_gbm_non_linear_all_quantiles_temp_33_40} presents the predictive accomplishment of the best-performing model—QR using GBM kernel—highlighting its capacity to capture the complex non-linear relationships at varying quantiles (0.05, 0.25, 0.50, 0.75, 0.95). From top to bottom, the figure sequentially exhibits the model's proficiency in forecasting HR at multiple quantiles, from the 5th to the 95th percentile. This projection is based on patient age, ranging from infancy to nearly 18 years (0-200 months), and BT between 34°C to 40.9°C. The x-axis denotes the patient's age in months, and the y-axis records HR in beats per minute (bpm). The subfigure presents the actual HR data as black dots, with the model's predicted values shown in red, facilitating a straightforward comparison of the observed and predicted HR. This allows for a direct appraisal of the model's precision in replicating the true HR values.

Furthermore, in response to the nature of the wide BT range (from 34°C to 40.9°C), Fig. \ref{fig:model_predictions} is introduced to deliver enhanced detail concerning the model's predictions, with respect to BT variations. Specifically, the experimental results for HR prediction for a specific quantile (0.05, 0.25, 0.50, 0.75, 0.95) are presented, respectively, in each row. A color gradient alongside each scatter plot translates the spectrum of colors to specific BT values, offering an instant visual guide. The scatter points, color-coded to represent varying BTs, effectively differentiate the HR predictions according to BT variations, thereby enriching the visual representation and enhancing the interpretability of the model's capacity to account for temperature-dependent HR variations. Consequently, analyzing the scatter plots across all the quantiles reveals a pattern: as BT increases, so does the predicted HR, while an inverse relationship is observed with increasing age, where the predicted HR generally declines. This visualization is not merely a reflection of the model's prediction accuracy; it also communicates critical insights for clinical applications, particularly as it highlights the expected HR values within the commonly encountered 5th to 95th percentile range in a clinical setting.

\begin{figure*}[h!]
    \centering
    
    \begin{minipage}{0.6\textwidth}
        \centering
        \includegraphics[width=\textwidth]{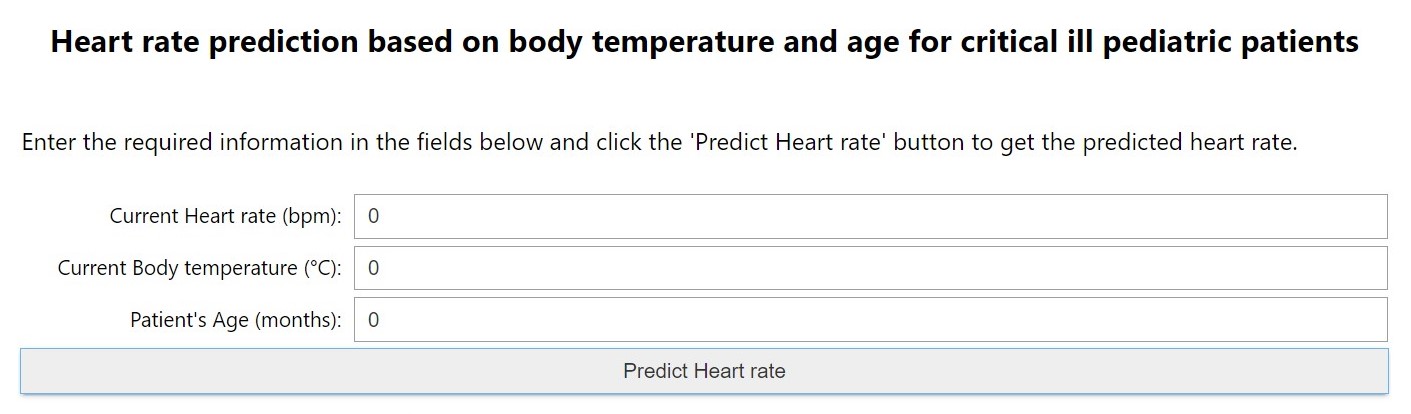}
        \caption*{(A)}
    \end{minipage}
            
    \begin{minipage}{0.35\textwidth}
        \centering
        \includegraphics[width=\textwidth]{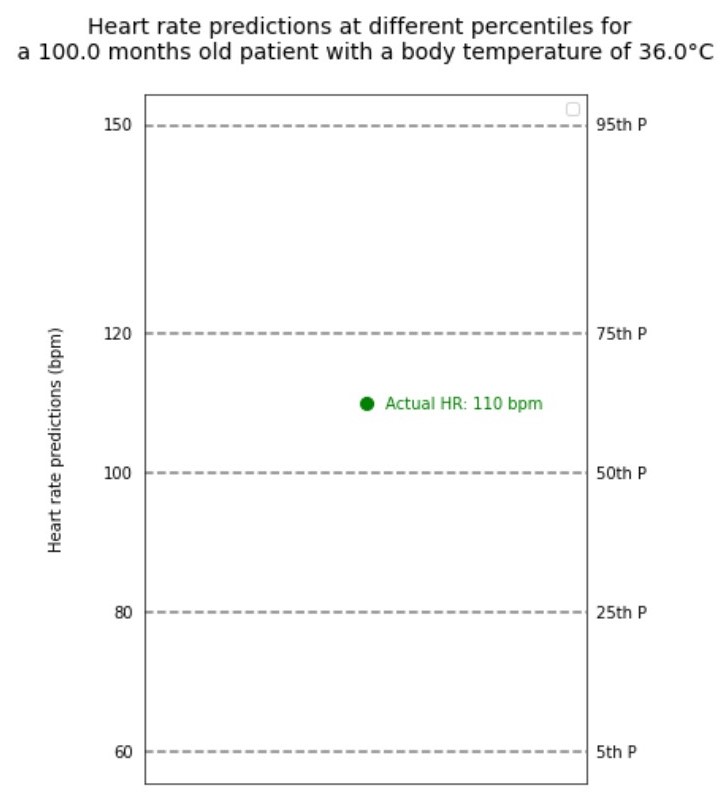}
        \caption*{(B)}
    \end{minipage}
    \hfill
    \begin{minipage}{0.35\textwidth}
        \centering
        \includegraphics[width=\textwidth]{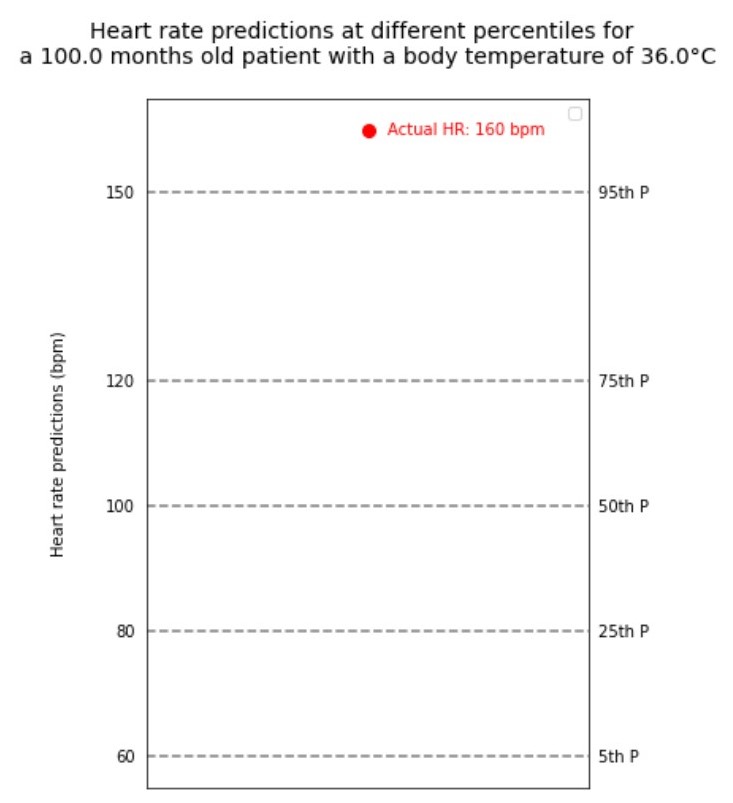}
        \caption*{(C)}
    \end{minipage}
    
    \caption{HR Predictions at different percentiles (5th, 25th, 50th, 75th, 95th) for a given patient's age and BT will be integrated into the CDSS system  (A) User interface featuring three key inputs - current HR, current BT, and patient's age - for HR predictions (B) Example of case scenario with a normal HR is indicated by a green dot, which falls within the normal range defined by the 5th and 95th percentiles  (C) Another example of a case scenario shows an abnormal HR, marked by a red dot located outside the normal range}
    \label{fig:ui_hr_prediction}
\end{figure*}

To enable clinicians to utilize our QR prediction model performed with GBM, we have developed an example of a simple user interface (UI). This UI processes three input parameters: the current HR, BT, and the patient's age. The output shows HR predictions across various percentiles (5th and 95th) tailored to a specific patient, as shown in Figure \ref{fig:ui_hr_prediction}. Importantly, our model defines a normal HR range within the 5th to 95th percentile. Within this range, a green dot signifies that the patient's HR falls within the expected parameters. On the other hand, a red dot serves as an immediate visual indicator of deviation from the norm, demanding prompt attention and intervention. As a result, this CDSS can determine in real-time whether the patient's HR is within the predicted range or needs more evaluation, thereby helping caregivers with clinical decision-making. 

This predictive model holds immense potential for integration within existing data representation developed by Yakob et al. (2024) for the CDSS in the PICU of CHUSJ \cite{yakob2024data}. However, it should be noted that integrating this tool in the PICU of CHUSJ does not require caregivers to manually enter patient information. In fact, the data representation for the CDSS consists of three levels: unit, patient, and system levels. The unit level allows visualization of all admitted patients, while the patient level enables assessment of the patient's status and progress \cite{yakob2024data}. Lastly, the system level presents many "groups of indicators related to human body systems and provides access to decision support tools developed for specific issues involving these systems \cite{yakob2024data}". Therefore, the patient's information is already integrated into the system automatically. Consequently, our predictive model will be generated and displayed at the system level of the CDSS.

Our model's limitation is that no HR prediction is provided if the inputs fall outside the ranges covered by our dataset. This underscores the importance of ensuring that the input parameters remain within the dataset's scope to obtain accurate predictions. Furthermore, clinicians must evaluate the predictions' interpretation and validation. This highlights the critical role of clinical expertise in assessing and confirming the accuracy and clinical relevance of the predicted HR values in this population.

\section{Conclusion}
In summary, this study represents a comprehensive and innovative approach to understanding the relationship between HR, BT, and age in children within the PICU. By employing a meticulous multi-stage data preprocessing strategy, the research aimed to reveal complex patterns that conventional models might overlook.

The findings of this study align with prior research by confirming the expected downtrend in HR with increasing patient age. However, the nuanced approach of grouping data by temperature rather than age allowed for a more granular exploration. Notably, the performance evaluation of various ML and DL algorithms yielded insightful results. Unlike some earlier studies that suggested linear correlations, conventional linear regression demonstrated limited effectiveness in capturing the linear relationship within the data. On the other hand, QR performed with advanced ML, such as the GBM model, exhibited superior performance, successfully uncovering non-linear relationships across a broad BT range from 33 to 40.9°C. Furthermore, the HR model predictions clearly show a downward HR trend with age and an upward trend with BT between the 5th and 95th percentiles. Based on that model, we created a simple user interface for caregivers. Based on age and BT, they can quickly determine in real-time whether a patient's HR falls within the normal range or not.

These findings have significant implications, highlighting the potential of ML and DL techniques to decode intricate associations in critically ill pediatric patients. The identified model enhances prediction capabilities and holds promise for early detection and developing more personalized and effective therapeutic interventions.

Looking forward, the paper suggests promising future directions and areas for improvement. One intriguing prospect involves observing a cohort of subjects upon whom clinicians have implemented interventions guided by the predicted HR model. These observations could involve observing trends and assessing the long-term effects of interventions. Extracting insights from the resulting conditions and outcomes would be interesting in refining future studies. Moreover, if the HR falls outside the 5th and 95th percentiles, the time spent in this range may be associated with bad outcomes. It would be valuable to investigate whether there are treatments that could reduce the duration of time spent in these extremes to enhance patient outcomes. Another direction for future research is to explore the interplay of variables, focusing on gender-based differences, considering that HR tends to be higher in women than men. Finally, incorporating other neural network architectures, such as Convolutional Neural Networks (CNN) or Transformers, could offer valuable insights.

\section*{Acknowledgment}

This work was supported in part by the Natural Sciences and Engineering Research Council (NSERC), in part by the Institut de Valorisation des données de l’Université de Montréal (IVADO), in part by the Fonds de la recherche en sante du Quebec (FRQS).

\bibliographystyle{IEEEtran}
\bibliography{IEEEabrv,Bibliography}
\begin{IEEEbiography}[{\includegraphics[width=1in, height=1.25in, clip, keepaspectratio]{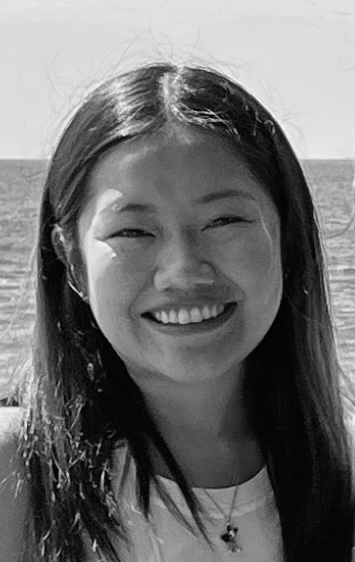}}]{Emilie Lu} received a B.Sc. degree in Biochemistry and Molecular Medicine from Montreal University (2017). Complementing this, she also earned a B.Eng. degree in Software Engineering from École de technologie supérieure (ÉTS) de Montreal (2022), and she’s currently completing a Master in Engineering in Healthcare Technology from ÉTS (2024). Her professional experience is marked by her contributions to medical imaging and surgical planning aspects. She has excelled in tasks such as segmentation for 3D modeling and the design of custom instruments, including implants for shoulder, hips, and knee procedures. Collaborating closely with surgeons, she actively engages in surgery planning, demonstrating her ability to integrate technology seamlessly into medical workflows. Her diverse skill set extends to her interests, encompassing machine learning, virtual and augmented reality, medical imaging, and orthopedics. Her passion for these fields reflects a commitment to staying at the forefront of technological advancements, specifically focusing on their applications in healthcare.
\end{IEEEbiography}

\begin{IEEEbiography}[{\includegraphics[width=1in, height=1.25in, clip, keepaspectratio]{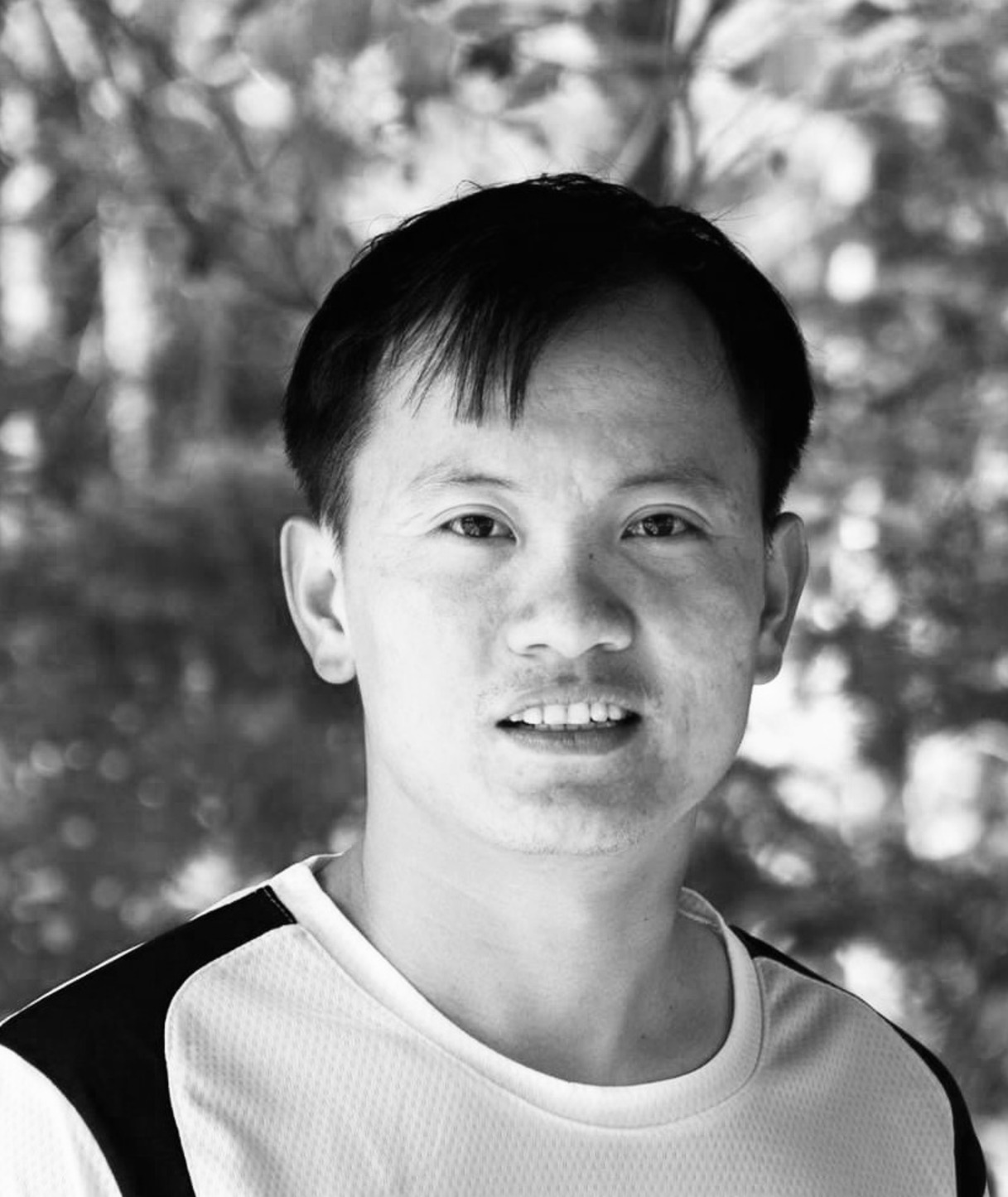}}]{Thanh-Dung Le} (Member, IEEE) received a B.Eng. degree in mechatronics engineering from Can Tho University, Vietnam, an M.Eng. degree in electrical engineering from Jeju National University, S. Korea, and a Ph.D. in biomedical engineering from \'{E}cole de Technologie Sup\'{e}rieure (ETS), Canada. He is a postdoctoral fellow at the Biomedical Information Processing Laboratory, ETS. His research interests include applied machine learning approaches for biomedical informatics problems. Before that, he joined the Institut National de la Recherche Scientifique, Canada, where he researched classification theory and machine learning with healthcare applications. He received the merit doctoral scholarship from Le Fonds de Recherche du Quebec Nature et Technologies. He also received the NSERC-PERSWADE fellowship in Canada and a graduate scholarship from the Korean National Research Foundation, S. Korea.
\end{IEEEbiography}

\begin{IEEEbiography}
[{\includegraphics[width=1in,height=1.25in, clip, keepaspectratio]{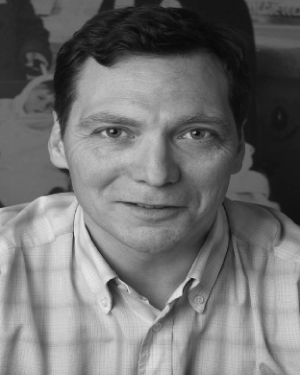}}]{Philippe Jouvet}received the M.D. degree from Paris V University, Paris, France, in 1989, the M.D. specialty in pediatrics and the M.D. subspecialty in intensive care from Paris V University, in 1989 and 1990, respectively, and the Ph.D. degree in pathophysiology of human nutrition and metabolism from Paris VII University, Paris, in 2001. He joined the Pediatric Intensive Care Unit of Sainte Justine Hospital—University of Montreal, Montreal, QC, Canada 2004. He is currently the Deputy Director of the Research Center and the Scientific Director of the Health Technology Assessment Unit at Sainte Justine Hospital–University of Montreal. He has a salary award for research from the Quebec Public Research Agency (FRQS). He currently conducts a research program on computerized decision support systems for health providers. His research program is supported by several grants from the Sainte-Justine Hospital, Quebec Ministry of Health, the FRQS, the Canadian Institutes of Health Research (CIHR), and the Natural Sciences and Engineering Research Council (NSERC). He has published more than 160 articles in peer-reviewed journals. Dr. Jouvet gave more than 120 lectures in national and international congresses.
\end{IEEEbiography}

\begin{IEEEbiography}[{\includegraphics[width=1in,height=1.25in, clip, keepaspectratio]{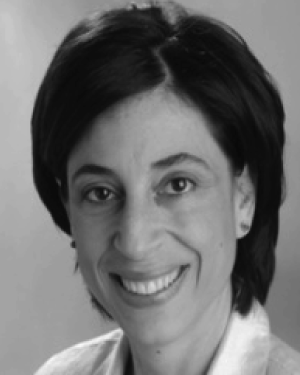}}]{Rita Noumeir} (Member, IEEE) received master's and Ph.D. degrees in biomedical engineering from École Polytechnique of Montreal. She is currently a Full Professor with the Department of Electrical Engineering, École de Technologie Superieure (ETS), Montreal. Her main research interest is in applying artificial intelligence methods to create decision support systems. She has extensively worked in healthcare information technology and image processing. She has also provided consulting services in large-scale software architecture, healthcare interoperability, workflow analysis, and technology assessment for several international software and medical companies, including Canada Health Infoway.
\end{IEEEbiography}

\end{document}